\documentclass[aps,prb,twocolumn]{revtex4}
\usepackage{graphicx}
\newcommand{\be}{\begin{equation}}
\newcommand{\ee}{\end{equation}}
\begin{document}
\title{Weak Localization in Graphene: Theory, Simulations and Experiments}
\author{M. Hilke, M. Massicotte, E. Whiteway, and V. Yu}
\affiliation{Department of Physics, McGill University, Montr\'eal, Canada H3A 2T8}
\date{\today}

\begin{abstract}
We provide a comprehensive picture of magnetotransport in graphene monolayers in the limit of non-quantizing magnetic fields. We discuss the effects of two carrier transport, weak localization, weak anti-localization, and strong localization for graphene devices of various mobilities, through theory, experiments and numerical simulations. In particular, we observe the weak localization of the localization length, which allows us to make the connection between weak and strong localization. It provides a unified framework for both localizations, which explains the observed experimental features. We compare these results to numerical simulation and find a remarkable agreement between theory, experiment and numerics. Various graphene devices were used in this study, including graphene on different substrates, such as glass and silicon, as well as low and high mobility devices.

\end{abstract}
\maketitle

\section{Introduction}

Graphene has attracted a considerable amount of attention due to the ease in isolating a single sheet of graphite via mechanical exfoliation.\cite{novo04,novo05} Despite the fact that it is only one atom thick, exfoliated graphene has shown extraordinary transport properties and can be used as a novel material for many potential applications. Its unique band structure has led to many interesting phenomena such as tunable charge carriers densities,\cite{Cooper12} anomalous integer quantum Hall effect \cite{novo04,zhang05} and ultrahigh mobilities at room temperature.\cite{bol08} The most noteworthy property of the band structure is the existence of two degenerate Dirac cones, \cite{Wallace47} which leads to two degenerate valleys (K and K'). The existence of these valleys, each with a linear dispersion, is the main reason that graphene has transport properties which differ from most other semiconductors or semimetals. At high magnetic fields, a striking example is the quantum Hall effect, where the Hall conductance quantization occurs in steps of 4$e^2/h$, because of the spin and valley degeneracy, and where the Landau level quantization is proportional to $\sqrt{B}$, where $B$ is the perpendicular magnetic field component. This square root dependence leads to a very large lowest Landau level splitting, where the field quantization has been observed up to room temperature.\cite{novo07} While the four-fold degeneracy can be lifted at very high fields in high mobility graphene, it is invisible at lower fields, which is the focus of this work.

The goal of this work is to present a comprehensive understanding of the magnetoresistance of monolayer graphene in a regime where the Landau quantization is not important, i.e., $\mu_qB\ll 1$, where $\mu_q$ is the quantum mobility. In this regime, the Landau quantization does not play any role, but the valley degeneracy, the linear dispersion and the underlying disorder potential lead to interesting magnetotransport phenomena such as two carrier transport (2CT), weak localization (WL), weak anti-localization (WAL), and strong localization (SL).

Most of the magnetotransport properties of graphene can be understood in terms of two types of scattering mechanisms: intervalley scattering, where electrons are scattered from one valley to the other (K to K') with a rate hereby noted as $\tau_e^{-1}$ and intravalley scattering, where electrons scatter within a valley as described by a rate $\tau_a^{-1}$. In general, intervalley stems from short range scattering, such as atomic defects, including grain boundaries, whereas intravalley scattering is long range and is typically stronger and also includes large scale inhomogeneities, as well as charged impurities in the substrate. Long range potential variations are responsible for the existence of both carriers types (electrons and holes) simultaneously at the charge neutrality point (CNP). The carrier distribution can be understood by assuming that the total carrier density is given by $n+p=\sqrt{n_0^2+(n-p)^2}$,\cite{Dorgan2010} where $n-p=C|V_G-V_D|/e$, $C$ is the gate capacitance, $V_G$ the gate voltage, and $V_D$ the gate voltage at the charge neutrality point (CNP) with a total residual density $n_0$. $n$ and $p$ are the carrier densities of electrons and holes, respectively. This allows us to define electron and hole mobilities, $\mu_n=\sigma_n/en$ and $\mu_p=\sigma_p/ep$, where $\sigma_n$ and $\sigma_p$ are the corresponding electron and hole conductivities. Experimentally, $\mu_n$ and $\mu_p$ are often slightly different, which is likely due to the asymmetry of the scattering potential for electrons and holes.

\section{Two carrier transport}

We start our magnetotransport analysis, by discussing the simplest non-trivial contribution, which is due to the simultaneous existence of the two types of carriers near the CNP. As long as $\mu_qB\ll 1$, this contribution can be evaluated using Drude's expression for two carriers, where $\rho_{xx}^{\eta}=\sigma_\eta^{-1}$ and $\rho_{xy}^\eta=\mbox{sign}(q_\eta)B/\eta$ for the two types of carrier densities ($\eta=n$ or $p$). The total resistivity is then given by $\rho_{tot}=(\hat{\sigma}_n+\hat{\sigma}_p)^{-1}$, where $\hat{\sigma}_\eta$ are the single band conductivity matrices. This yields the following field dependence of the magnetoresistivity

\begin{equation}
\rho_{xx}^{tot}=\frac{(np)^2(\sigma_n+\sigma_p)+B^2\sigma_n\sigma_p(n^2\sigma_p+p^2\sigma_n)}
{(np)^2(\sigma_n+\sigma_p)^2+B^2(\sigma_n\sigma_p)^2(n-p)^2}.
\end{equation}
From this it follows that the relative field dependence can be written as
\begin{equation}
\frac{\Delta\rho_{xx}}{\rho_{xx}}=\frac{\rho_{xx}(B)-\rho_{xx}(0)}{\rho_{xx}(0)}=\frac{(B/B_0)^2}{1+(B/B_1)^2},
\label{2CT}
\end{equation}
\noindent where
\begin{equation}
B_0=\frac{enp(\sigma_n+\sigma_p)}{\sqrt{\sigma_n\sigma_p}(p\sigma_n+n\sigma_p)}\mbox{ \& }B_1=\frac{enp(\sigma_n+\sigma_p)}{|n-p|\sigma_n\sigma_p}.
\label{B0B1}
\end{equation}

Expressing the conductivities in equation (\ref{B0B1}) in terms of mobilities, equating the zero field resistivity to $\rho_{xx}=(\sigma_n+\sigma_p)^{-1}$ and identifying $\rho_{max}$ as the zero field resistivity at the CNP, we obtain

\begin{equation}
B_0=\frac{\rho_{max}}{\sqrt{\mu_n\mu_p}\rho_{xx}}\simeq\frac{en_0\rho_{max}^2}{\rho_{xx}} \mbox{ \& }B_1\simeq \frac{n_0B_0}{|n-p|}.
\end{equation}

Hence, as a function of $B$ the 2CT gives rise to a parabolic positive relative magnetoresistance before saturating to $n_0/|n-p|$ when $B\simeq B_1$, which is valid as long as $\mu_qB\ll 1$. Close to the CNP, i.e., $|n-p|\ll n_0$, $B_0$ is simply equal to the inverse median mobility and $B_1\gg B_0$, which is the regime where the 2CT effect is the largest.  Away from the CNP ($|n-p|\gg n_0$), we have a maximum $\Delta \rho/\rho\simeq n_0/n_H$, which decreases away from the CNP, since  $n_H\simeq |n-p|$, which is the Hall density, increases. The 2CT will play an important role, as long as $\mu_qB\ll 1$, to give rise to a positive magnetoresistance due to large scale inhomogeneities. This effect is very important in graphene as compared to other two dimensional systems because of the absence of a gap between the electron and hole carriers. The 2CT will serve as basis for the understanding of the classical contribution due to long range scattering, which will be present at all temperatures and even increases with temperature, since $n_0$ increases with temperature due to the thermal activation of the electron and hole carriers.

\section{Theory of weak localization}

Moving beyond the Drude and classical description of transport, we need to include quantum effects, of which coherent backscattering is the most important contribution. Coherent backscattering leads to WL.\cite{Hikami80,Altshuler80} To obtain an expression for the WL correction, we have to evaluate the return probability of all possible trajectories.\cite{Chacra86} At zero magnetic field, the coherent return probability $P_{ret}$ can be expressed as:\cite{Beenakker88}

\begin{eqnarray}
P_{ret} &=& \int_{0}^{\infty}\frac{1}{4\pi Dt} (1-e^{-t/\tau_e})e^{-t/\tau_\phi}dt
\nonumber\\
&=& \frac{1}{4\pi D}\ln\left(\frac{\tau_\phi}{\tau_e}+1\right),
\label{integral}
\end{eqnarray}

\noindent where the first term $(4\pi Dt)^{-1}$ in the integrant is the return probability ($P_0(t)$) for diffusion constant $D$, the second term represents the short time cut-off ($\tau_e$), below which no elastic scattering occurs, and the third term is the phase coherence time ($\tau_\phi$) cut-off, beyond which phase coherence is lost. At low temperatures, where $\tau_\phi\gg\tau_e$, this leads to the typical logarithmic WL correction of the conductivity $\delta\sigma=-\frac{4e^2D}{h}P_{ret}\simeq -\frac{e^2}{\pi h}\ln(\tau_\phi/\tau_e)$.

To evaluate the magnetic field dependence of the return probability $P_B(t)$, we have to solve for the return probability according to the field dependent diffusion equation:\cite{Chacra86}

\begin{equation}
[\frac{\partial}{\partial t}+D(i\nabla+\frac{2e}{\hbar}A)^2]P_B(r,r',t)=\delta(r-r')\delta(t),
\end{equation}

\noindent where $A$ is the vector potential. The solution to this diffusion equation can be evaluated for $r=r'$ and is given by

\begin{equation}
P_B(t)=\frac{eB/h}{\sinh (t/2\tau_B)},
\end{equation}

\noindent where $\tau_B=\hbar/4eBD$. At zero field we recover $P_0(t)$. In the long time limit, where we have $P_B(t)\rightarrow (eB/h)e^{-t/2\tau_B}$, this leads to the field induced destruction of the phase coherence when approximately one flux penetrates the area $L^2_B=D\tau_B$. The full magnetic field dependence is obtained by inserting $P_B(t)$ into the expression of the coherent return probability and evaluating the integral, i.e.,

\begin{eqnarray}
P_{ret} &=& \int_{0}^{\infty}P_B(t) (1-e^{-t/\tau_e})e^{-t/\tau_\phi}dt
\\
&=& \frac{1}{4\pi D}\left[\psi\left(\frac{1}2+\frac{B_e+B_\phi}{B}\right)-\psi\left(\frac{1}2+\frac{B_\phi}{B}\right)\right],\nonumber
\label{integral}
\end{eqnarray}

\noindent where $B_\phi=\hbar/4eD\tau_\phi$ and $B_e=\hbar/4eD\tau_e$ are the characteristic magnetic fields for $\tau_\phi$ and $\tau_e$, whereas $\psi$ is the digamma function. The magnetic field correction to the coherent return probability, is then given by

\begin{eqnarray}
\Delta P_{ret}&=&P_{ret}(B)-P_{ret}(0)\nonumber\\
&=& \frac{1}{4\pi D}\left[F\left(\frac{B}{B_\phi+B_e}\right)-F\left(\frac{B}{B_\phi}\right)\right],
\label{DeltaPret}
\end{eqnarray}

\noindent where $F(z)=\psi(1/2+1/z)+\ln(z)$, $F(z)\simeq z^2/24$ for $z\ll 1$, $F(z)\simeq\ln(1+z/4e^\gamma)$ for $z\gg 1$, and $\gamma$ is the Euler constant. The second term in equation (\ref{DeltaPret}) is responsible for the reduction of the return probability due to the presence of a magnetic field, which leads to the observed peak in resistance of width $\sim B_\phi$ around zero magnetic field, characteristic of weak localization in disordered systems.\cite{Bergmann84} The first term renormalizes the coherent return probability at large fields, i.e., $4\pi D\Delta P_{ret}(B\rightarrow\infty)=\ln(B_\phi/(B_\phi+B_e))$. This expression leads to the usual WL correction to the conductivity when expressing the conductivity correction in terms of the coherent return probability correction, i.e., $\Delta \sigma=-4e^2D\Delta P_{ret}/h$.\cite{Chacra86,WLth}

In deriving equation (\ref{DeltaPret}) we assumed the existence of only two types of scattering mechanisms $\tau_\phi$ (phase breaking) and $\tau_e$ (elastic). However, in graphene, we have two very different types of elastic scattering mechanisms: those that scatter electrons within a valley (intravalley, $\tau_a$) and those that scatter between valleys (intervalley, K to K', $\tau_e$). For coherent backscattering, only intervalley scattering contributes, which also includes inelastic scattering ($\tau_\phi$). Hence, the short time cutoff scattering time is provided by $\tau_e$ in graphene, which is short ranged and can be due to grain boundaries or lattice defects. However, intravalley scattering is usually much stronger in graphene, but doesn't contribute to coherent backscattering and can lead to enhanced forward scattering at zero magnetic field and positive magnetoresistance.

In an important work by McCann and co-workers,\cite{WLth} based on earlier work in honeycomb lattices,\cite{ando02} the authors have obtained a general expression for the WL and WAL correction specific to graphene, which determines the dependence of the magnetoresistivity as a function of $B$ involving the scattering parameters $\tau_a$ and $\tau_e$ explicitly. They obtained the following expression:

\begin{equation}
\frac{\pi h}{e^2}\Delta \sigma = F\left(\frac{B}{B_\phi}\right)-F\left(\frac{B}{B_\phi+B_e}\right)-2F\left(\frac{B}{B_\phi+B_\star}\right),
\label{McCann}
\end{equation}

\noindent where $B_\star=B_a+B_e/2$. This expression can also be obtained directly using equation (\ref{integral}) by replacing the short time cut-off $e^{-t/\tau_e}$ in the magnetic field correction by $(e^{-t/\tau_e}+2e^{-t/\tau_\star})$. While the first two terms in equation (\ref{McCann}) lead to the usual WL localization effect as in equation (\ref{DeltaPret}), the last term leads to WAL due to the presence of intravalley scattering as illustrated in figure 1. This term becomes important when $B$ approaches $B_\star$ and leads to a positive magnetoresistance contribution.

\begin{figure}[ptb]
\begin{center}
\vspace{0cm}
\includegraphics[width=3.5in]{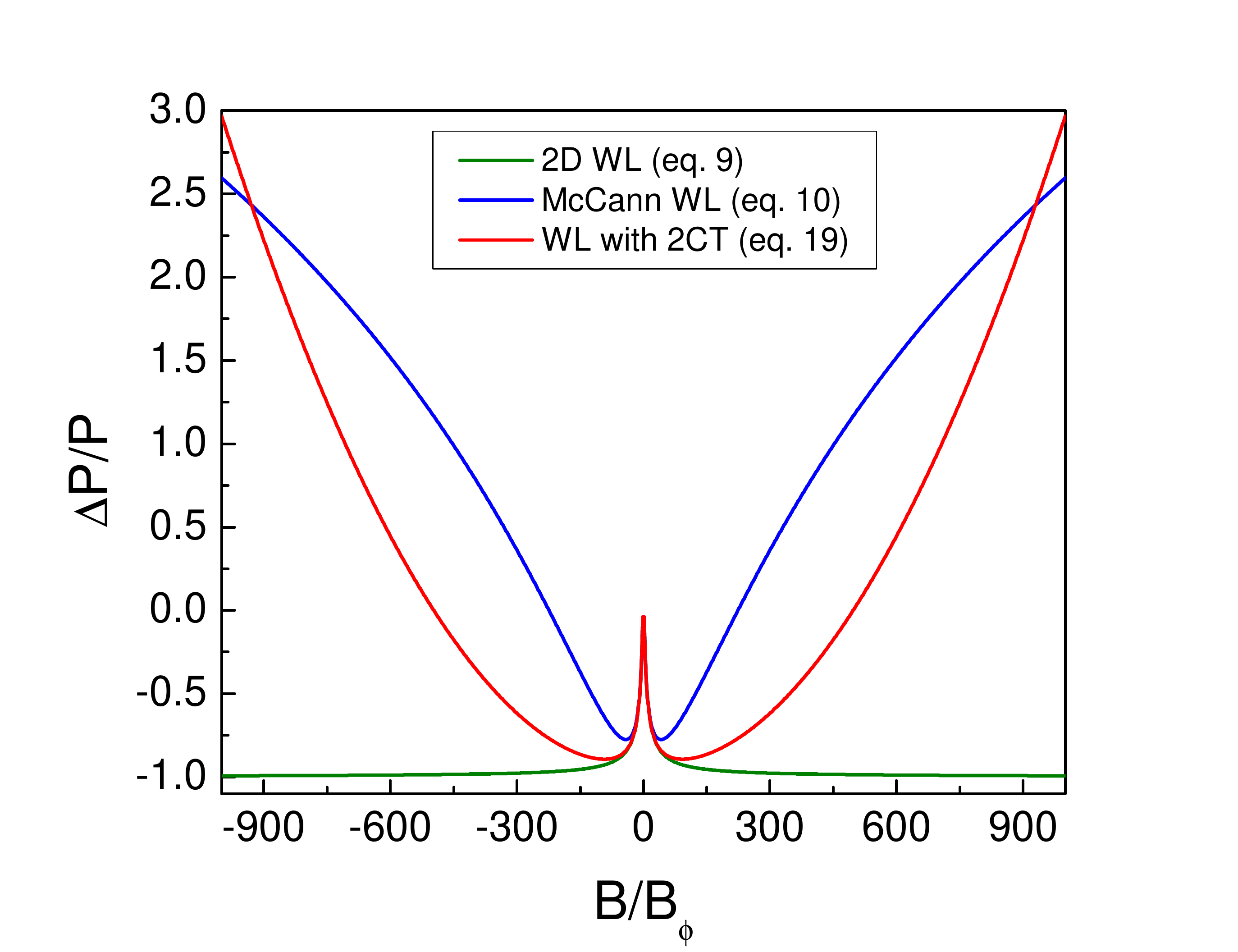}
\vspace{0cm}
\caption{The relative return probability is shown as a function of the magnetic field for the different models. The green curve is from expression (\ref{DeltaPret}), the blue curve is from expression (\ref{McCann}) with $B_\star=50B_\phi$, and the red curve is obtained from expression (\ref{expfit}) with $B_0=500B_\phi$ and $B_1=\infty$. For all curves we assumed $B_e=B_\phi$. }%
\label{localization}%
\end{center}
\end{figure}

\section{Strong localization}

In some graphene samples disorder can be very important, like for instance in intentionally disordered exfoliated graphene, which was shown to lead to strong localization.\cite{moser10}
Strong localization or Anderson localization \cite{Anderson58} is obtained when the transmission is exponentially suppressed, i.e., when a coherent wave is backscattered due to wave interference. Hence, strong localization is a direct consequence of coherent backscattering. In the language of Anderson localization, the important parameter is the localization length $L_c$, which measures the exponential increase of the resistance with size. In a system where the coherence length is infinite, the inverse localization length can be defined by

\be
L_c^{-1}=\frac{1}{L}\lim_{L\rightarrow\infty} \ln(R(L)) ,
\ee
where $L$ is the length of the system and $R$ the resistance. In order to incorporate strong localization in our discussion on weak localization, we can define the localization time as $\tau_c=L^2_c/D$, which represents the time before the charge carrier is localized in a diffusive system. The coherent return probability then has to be modified according to

\begin{eqnarray}
P_{ret} &=& \int_{0}^{\infty}P_B(t)e^{-t/\tau_c} (1-e^{-t/\tau_e})e^{-t/\tau_\phi}dt
\\
&=& \frac{1}{4\pi D}\left[\psi\left(\frac{1}2+\frac{B_e+B_{c\phi}}{B}\right)-\psi\left(\frac{1}2+\frac{B_{c\phi}}{B}\right)\right],\nonumber
\label{integral2}
\end{eqnarray}

\noindent where $B_c=\hbar/4eD\tau_c=\hbar/4eL^2_c$ and $B_{c\phi}=B_c+B_\phi$. The effect of strong localization, is therefore simply to replace $B_\phi$ by $B_{c\phi}$. This is very intuitive, since if $L_c\ll L_\phi$, $L_c$ will play the role of $L_\phi$ in determining the magnetic field correction, since no closed trajectory can occur beyond $L_c$. The relative change in coherent return probability can then be written as

\begin{equation}
\frac{\Delta P_{ret}}{P_{ret}}= \frac{F\left(\frac{B}{B_e+B_{c\phi}}\right)-F\left(\frac{B}{B_{c\phi}}\right)}{\ln(B_e+B_{c\phi})-\ln(B_{c\phi})},
\label{StrongL}
\end{equation}

\noindent where the field dependence is mainly determined by $B_{c\phi}$.  Since we define the  localization length assuming a coherent system, the field induced relative change in $L_c$ is directly determined by the change in the relative coherent return probability for zero dephasing:

\begin{equation}
\frac{\Delta L_c}{L_c}\simeq \left.-\frac{\Delta P_{ret}}{P_{ret}}\right|_{B_\phi=0}.
\label{LcandPeq}
\end{equation}

Equation (\ref{LcandPeq}) tells us how strong localization is connected to the coherent return probability and therefore to WL. Since it is quite challenging to obtain an expression for the localization length in two dimensions directly, we will resort to evaluating the localization length numerically below.

\section{Graphene nanoribbons}
In many situations, both experimentally and numerically, one deals with a situation where the sample width is finite. This leads to a different field dependence, since the sample effectively becomes quasi-one dimensional (Q1D) instead of 2D,  when $L_\phi> W$ and $W$ is the width of the graphene ribbon. In this limit, the governing diffusion equation is no longer 2D, but has to be replaced by the 1D analogue, i.e.,  $P_B(t)=(1/\sqrt{4\pi Dt})e^{-t/\tau_{b}}$,\cite{Beenakker88} where the first term represents 1D diffusion at zero field and the second term represents the field induced destruction of phase coherence in Q1D. $\tau_{b}$ is given as usual by $L_{b}^2/D$, where $BWL_{b}=\sqrt{3/4}\hbar/e$ scales with the flux quantum through the relevant characteristic area $WL_{b}$.\cite{WLth} This leads to

\begin{eqnarray}
P_{ret} &=& \int_{0}^{\infty}\frac{e^{-t/\tau_{b}}}{\sqrt{4\pi Dt}} (1-e^{-t/\tau_e})e^{-t/\tau_\phi}dt
\\
&=& \frac{1}{2 \sqrt{D}}\left[\frac{1}{\sqrt{\tau_\phi^{-1}+\tau_{b}^{-1}}}-\frac{1}{\sqrt{\tau_\phi^{-1}+\tau_{b}^{-1}+\tau_e^{-1}}}\right].\nonumber
\label{integral3}
\end{eqnarray}

When $\tau_e^{-1}\gg\tau_\phi^{-1}+\tau_{b}^{-1}$ we obtain the following expression for the magnetic field correction of the coherent return probability:

\begin{equation}
\Delta P_{ret}=\frac{L_\phi}{2D} \left[1/\sqrt{1+(4/3)(eBWL_\phi/\hbar)^2}-1\right].
\label{ClassicWL}
\end{equation}

Recalling that $\Delta \sigma=-4e^2D\Delta P_{ret}/h$, we recover the expression for the WL correction to the conductivity obtained by McCann and co-workers.\cite{WLth}

For the strong disorder case, we have to again introduce the localization time $\tau_c$, which has the effect of simply replacing $1/L_\phi^2$ by $1/L_{c\phi}^2=1/L_\phi^2+1/L_c^2$. We then obtain for the relative change in the coherent return probability,

\begin{equation}
\frac{\Delta P_{ret}}{P_{ret}}= \left[1/\sqrt{1+B^2/12B_{c\phi}B_W}-1\right],
\label{Q1D}
\end{equation}

\noindent where $B_W=\hbar/4eW^2$. This is the result for the Q1D limit, when $L_\phi$ or $L_c$ $\gg W$. Here again we have $\Delta L_c/L_c\simeq \left.-\Delta P_{ret}/P_{ret}\right|_{B_\phi=0}$ for the relative change of the localization length.

\section{Numerical simulations}
We now move to evaluate $L_c$ numerically for graphene assuming $L_\phi\gg L_c,W$. There are two different limits: 2D and Q1D, i.e., $L_c \ll W$ (the high disorder limit) and $L_c \gg W$ (the low disorder limit), respectively. These two limits are mainly determined by the amount of disorder in the system. For the numerical simulations, we will only consider short ranged disorder, since this is the relevant source for WL. The effect of long range disorder is to induce the co-existence of electrons and holes close to the charge neutrality point as well as WAL.

To model the system accurately, we consider a tight binding model with a honeycomb structure with two types of edges: armchair and zigzag. The hopping term is given by $t\simeq$3 eV. The two ends of the graphene device are assumed connected to wide ohmic contacts with a quadratic dispersion. The short range disorder is implemented by adding a random onsite potential $v_i$ on every atom site uniformly distributed with $-V/2<v_i<V/2$. The magnetic field is simply a phase factor $\phi$ in the hopping term, where $2\pi\phi=1$ corresponds to one flux quantum in a hexagon. The two terminal transmission probability $T$ is then evaluated iteratively as a function of the length, $L$, of the system, using an efficient iterative Green's function method.\cite{Rotter00} The average two terminal conductance $\langle T(L) \rangle e^2/h$, resistance $\langle 1/T(L) \rangle h/e^2$ or logarithm conductance $\langle \ln (T(L)) \rangle$ are obtained by averaging over many disorder configurations ($\langle\cdot\rangle$). For a given $W$ and at large enough $L$, the system is always localized due to Anderson localization and $L_c$ can be extracted whenever $L\gg L_c$.

\begin{figure}[ptb]
\begin{center}
\vspace{0cm}
\includegraphics[width=3.5in]{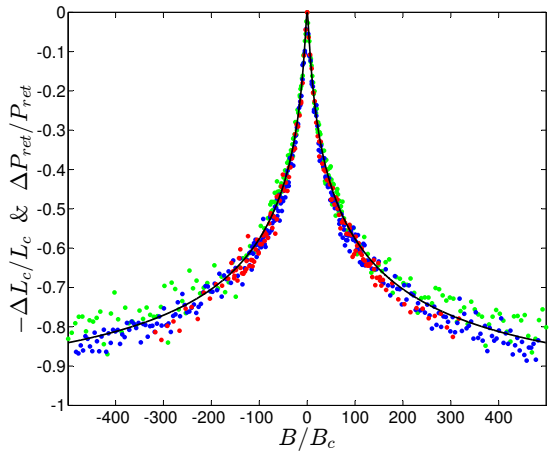}
\vspace{0cm}
\caption{We compare the analytical expression (solid line) for $\Delta P_{ret}/P_{ret}$ from equation (\ref{StrongL}) with the numerically obtained relative change in the localization length using the parameter $B_e=100B_c$. The different dot colors (green, blue, red) represent different widths (40$\times a$, 80$\times a$, 120$\times a$) of the simulated graphene device, where $a$ is the lattice constant.}%
\label{LcandP}%
\end{center}
\end{figure}

We now move to compare the numerically obtained $L_c$ with the analytical expressions derived above for the relative change in $L_c$, i.e., testing equation (\ref{LcandPeq}). In the 2D limit, where $W\gg L_c$, the field dependence of the coherent return probability is determined by equation (\ref{StrongL}), where $B_{c\phi}=B_c$ in the coherent limit. The numerical field dependence of the relative $L_c$ is then rescaled by $B/B_c$, where $B_c\sim L_c^{-2}$ and can be compared to the relative coherent return probabilities as shown in figure \ref{LcandP}. The agreement is quite remarkable with only two fitting parameters $B_e$ and $B_c$. Away from the Dirac point we expect to have $B_e/B_c\simeq W$, where $W$ is approximately the number of quantum channels in a quasi one dimensional system. We indeed obtained $B_e/B_c\simeq 100$ for the comparison to the numerical data where $40<W<120$ in units of the lattice constant.

In the Q1D limit, where $W\ll L_c$, the field dependence is rescaled by $B/\sqrt{12B_WB_c}$, where $\sqrt{B_WB_c}\sim 1/WL_c$ as determined from equation (\ref{Q1D}).

\begin{figure}[ptb]
\begin{center}
\vspace{0cm}
\includegraphics[width=3.5in]{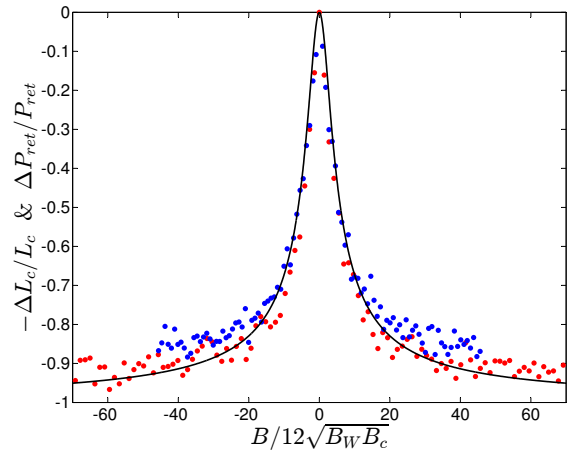}
\vspace{0cm}
\caption{We compare the analytical expression (solid line) for $\Delta P_{ret}/P_{ret}$ from equation (\ref{Q1D}), relevant to the quasi 1D situation, with the numerically obtained relative change in the localization length using the parameter $B_e=100B_c$. The different dot colors (blue, red) represent different widths (80$\times a$, 120$\times a$) of the simulated graphene device, where $a$ is the lattice constant.}%
\label{LcandP1D}%
\end{center}
\end{figure}

In both limits, the agreement between the numerically determined relative localization length and the relative coherent return probability is excellent, which confirms that indeed $\Delta L_c/L_c\simeq \left.-\Delta P_{ret}/P_{ret}\right|_{B_\phi=0}$. Hence, the field dependence of the coherent return probability determines $B_c$, which in turn determines $L_c$ and therefore provides us with a way to extract the localization length simply from the magnetic field dependence.

It is now possible to make the connection between $L_c$ and the resistance, since $\ln(R_{xx})\rightarrow L/L_c$. Therefore, small variations in $L_c$ lead to $\Delta R_{xx}/R_{xx}\simeq-L\Delta L_c/L_c^2$. This is assuming that $L_\phi\gg L$. However, in experiments, where this is often not the case, then $L_\phi$ plays the role of effective sample size ($L_\phi =L$) so that when $L_\phi\ll L$ we can write $\Delta R_{xx}/R_{xx}\simeq-L_\phi\Delta L_c/L_c^2$ instead. This is often used in experiments to extract $L_c$ from the temperature dependence of the resistance, since $L_\phi$ also depends on temperature.\cite{zhu11} Here, we restrict ourselves to a fixed temperature, where $L_\phi$ is constant, but $L_c$ is magnetic field dependent. Hence, when $L_\phi\ll L$ we have using equation (\ref{LcandPeq})

\begin{equation}
\left.\frac{\Delta R_{xx}}{R_{xx}}\right|_{ret}\simeq\frac{\Delta P_{ret}}{P_{ret}}\cdot \frac{L_\phi}{L_c}.
\label{linapprox}
\end{equation}

For small variations ($\Delta R/R\ll 1$), different contributions to the resistance are additive, hence the total effect on the resistance, including the two carrier effect, yields

\begin{equation}
\left.\frac{\Delta R_{xx}}{R_{xx}}\right|_{tot}\simeq \frac{L_\phi}{L_c}\frac{F\left(\frac{B}{B_e+B_{c\phi}}\right)-F\left(\frac{B}{B_{c\phi}}\right)}{\ln(1+B_e/B_{c\phi})}+\frac{(B/B_0)^2}{1+(B/B_1)^2},
\label{expfit}
\end{equation}

\noindent which allows the extraction of all relevant length scales from the magnetic field dependence alone, recalling that $L_c^{-2}+L_\phi^{-2}=4eB_{c\phi}/\hbar$, $L_e^{-2}=4eB_{e}/\hbar$, $B_0\simeq\rho_{max}/\mu\rho_{xx}$, and $B_1\simeq B_0n_0/n_H$.

\section{Experiments:}

While WL was observed in graphene in a number of previous experiments,\cite{morozov08,tikh08,ki08,mason10} and fitted using McCann's WL expression,\cite{WLth} they were generally limited to low disorder. Here we will use expression (\ref{expfit}) derived above and valid for all disorder strengths and fit it to the experimental data. We performed experiments on large scale graphene as well as lithographically defined Hall bars and graphene nano-ribbons, in addition to large (over 100 $\mu$m) single crystal graphene. Monolayers of graphene were grown by chemical vapor deposition (CVD) of hydrocarbons on 25 $\mu$m-thick commercial Cu foils. The CVD process used was similar to those described in previous works\cite{reina08,li09,Whiteway10,Yu11}.

\begin{figure}[!h]
\includegraphics[width=3.5in]{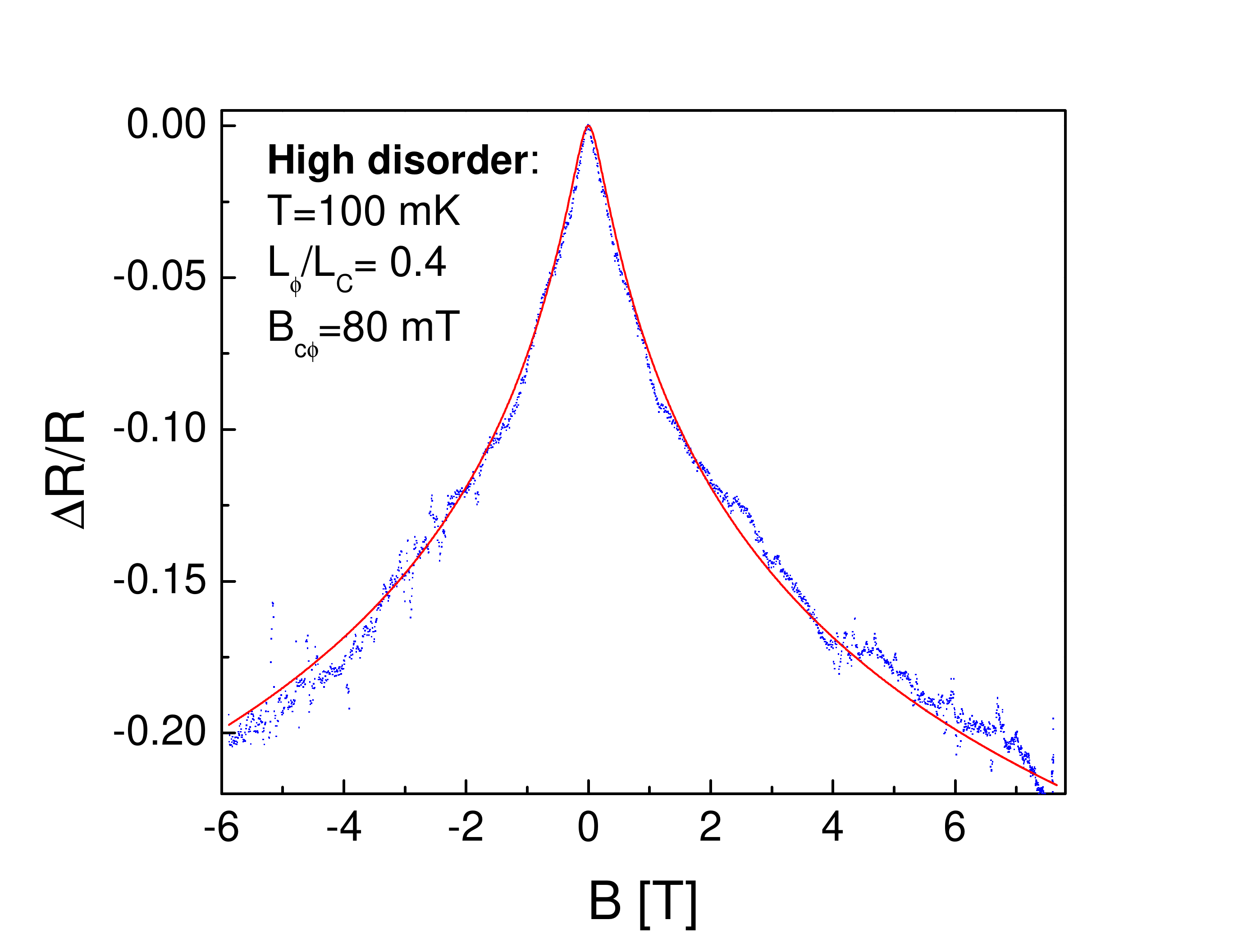}
\caption{The measured magneto-resistance of a low mobility ($\mu <100$cm$^2$/Vs) graphene Hall bar device at 100mK. Shown is the relative resistance (blues dots) and the fit using equation (\ref{expfit}) with fitting parameters $B_{c\phi}=80$mT, $B_e=10$T and $L_\phi/L_c=0.4$ (red line). $B_0$ and $B_1$ are larger than 100T and do not affect the fit.}
\label{lowmu}
\end{figure}

We start with low mobility samples, which cannot be fitted with the standard WL theory \cite{WLth} alone because of the importance of strong localization in this regime. Instead we use the expression derived in equation (\ref{StrongL}), which incorporates the effect of strong localization. In figure \ref{lowmu} we show the relative resistance change of a representative low mobility graphene Hall bar. Typical features of low mobility samples are a wide peak around zero magnetic field and an important relative resistance change. Using the best fit over the entire magnetic field range, we obtain $L_\phi\simeq 50$nm and $L_c\simeq 120$nm, which shows that the localization length is of the same order of magnitude as the phase coherence length. The agreement between the fit and the data is quite remarkable over the entire available magnetic field range and only relies on three fitting parameters, in this case $B_{c\phi}$, $B_e$ and $L_\phi/L_c$. The other parameters $B_0$ and $B_1$ are too large to significantly affect the fit. For even higher disorder, we expect $L_c$ to be smaller than $L_\phi$ and we would need to modify our assumption of the linear approximation used in equation (\ref{linapprox}). Instead, we expect the change in magnetoresistance to depend exponentially on $L_c(B)$ and its associated field dependence. This is amplified further for long but narrow samples (nanoribbons).

\begin{figure}[!h]
\includegraphics[width=3.5in]{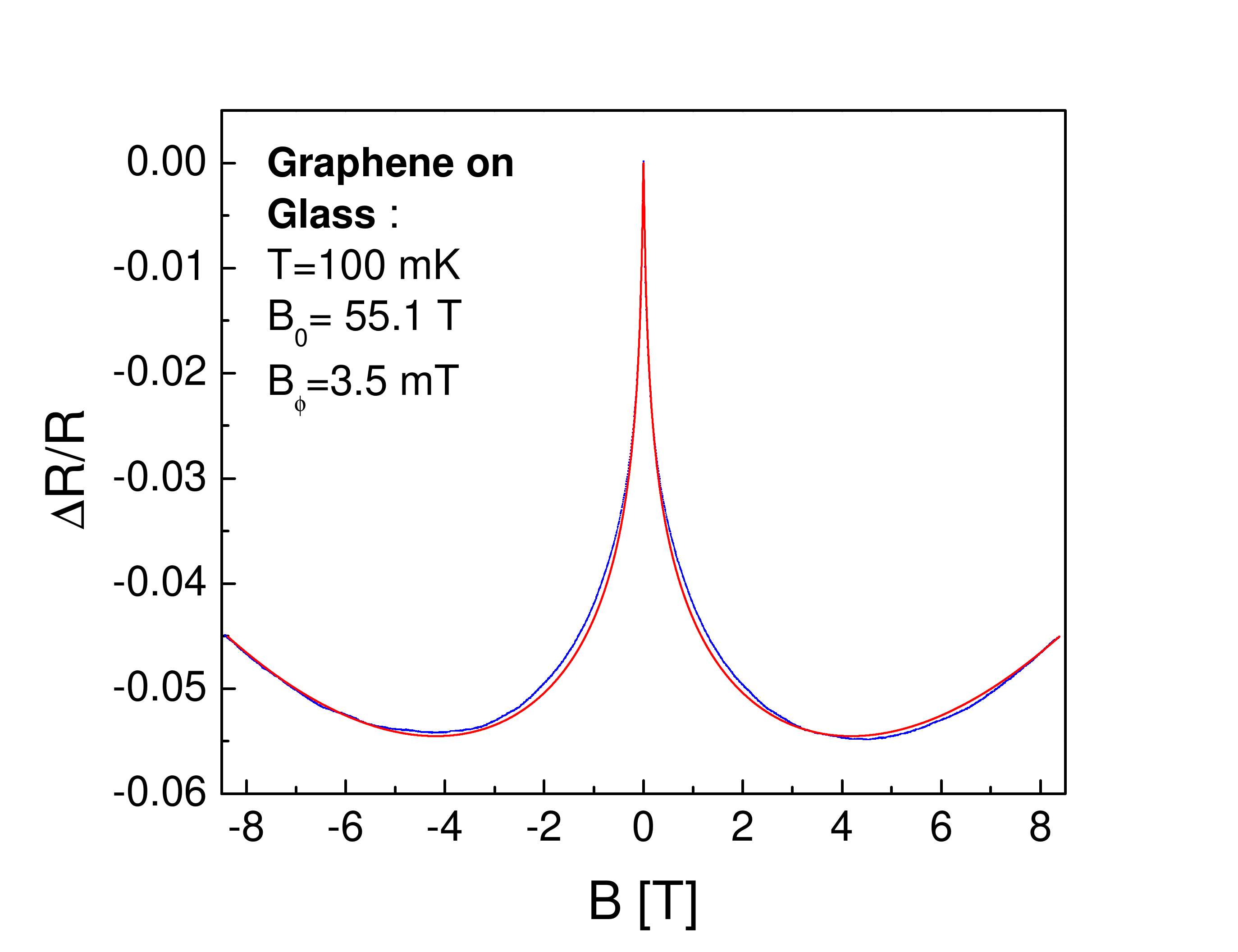}
\caption{The measured magneto-resistance of a large scale graphene sample on glass (blue dots). Shown is the relative resistance and the fit (red line) using equation (\ref{expfit}) with fitting parameters $B_{c\phi}=3.5$mT, $L_\phi/L_c=0.0925$, $B_0=55.1T$ and $B_e=9$T.}
\label{Glass}
\end{figure}

The next step is to look at a large scale ($\sim$cm) graphene sample grown by CVD deposited on a glass slide with evaporated gold as contacts. The magnetic field dependence is shown in figure \ref{Glass}, which shows a relatively broad peak at zero magnetic field. There is also a parabolic increase in resistance at large field, which we attribute to 2CT.  Using equation (\ref{expfit}) to fit the data we extract the following length scales $L_\phi\simeq 220$nm, $L_c\simeq 2.3\mu$m, and $L_e\simeq 4.3$nm. Here again, the agreement between the fit and the data is quite remarkable, considering that the fit extends over the entire magnetic field range.

\begin{figure}[!h]
\includegraphics[width=3.5in]{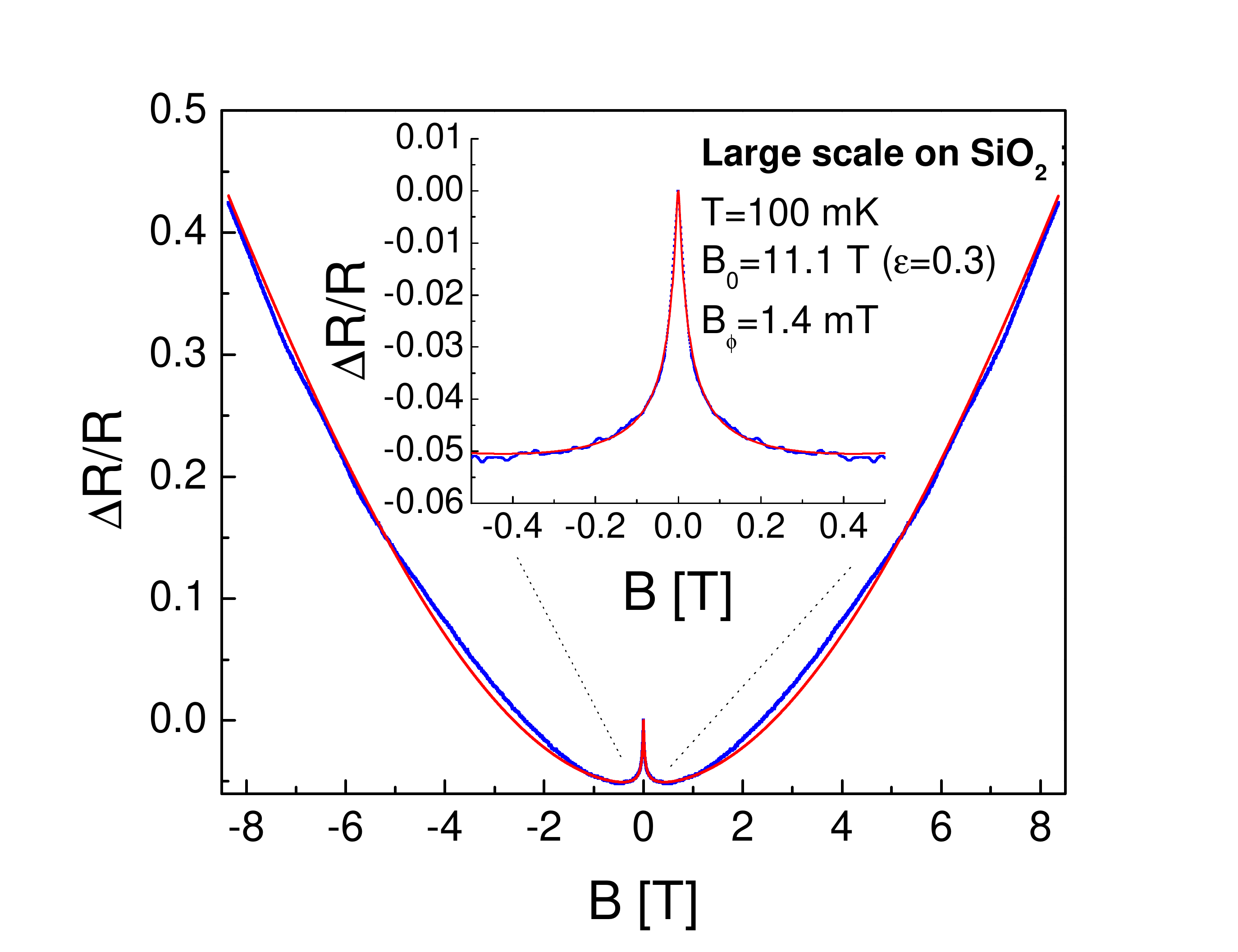}
\caption{The measured magneto-resistance of a large scale graphene sample on SiO$_2$/Si (blue dots). Shown is the symmetrized ($R_{xx}(B)+R_{xx}(-B)$) relative resistance and the fit (red line) using equation (\ref{expfit}) with fitting parameters $B_{c\phi}=1.4$mT, $L_\phi/L_c=0.055$, $B_0=11.1T$, $B_e=11$mT and $B_1=20.3T$ and a zoom-in of the low field part in the inset.}
\label{C13}
\end{figure}

The next sample we are considering is a large scale ($\sim$cm) sized graphene sample grown by CVD and deposited on a standard SiO$_2$/Si substrate, where the doped silicon can be used as a back gate. The main difference is that this sample was grown using isotopically pure C$^{13}$ methane gas instead of C$^{12}$, so that the Raman peaks are shifted by $\sqrt{12/13}$. \cite{Bernard11} However, the different isotope does not affect the magnetotransport. The magnetic field dependence is shown in figure \ref{C13}, which shows a narrow weak localization peak at zero magnetic field. There is also a large parabolic increase in resistance at large field, which we attribute again to the 2CT. Using equation (\ref{expfit}) to fit the experimental data we obtain $L_\phi\simeq 330$nm, $L_c\simeq 6\mu$m, $L_e\simeq 120$nm and a mean field effect mobility of $\mu\simeq$0.18m$^2$/Vs, which is common for large scale CVD graphene, and a residual density of $n_0\simeq 7.0\times 10^{11}$cm$^{-2}$.

Interestingly, this is a prime example of a large parabolic-like positive magneto-resistance background, which cannot be fitted using WAL as per equation (\ref{McCann}), since WAL gives a negatively curved magneto-resistance at large fields. Here, we clearly demonstrate that the origin of the large positive magnetoresistance background is likely due to the 2CT because of the coexistence of both n and p type carriers. Strikingly, the quality of the fit is very remarkable over the entire available experimental magnetic field range.

\begin{figure}[!h]
\includegraphics[width=3.5in]{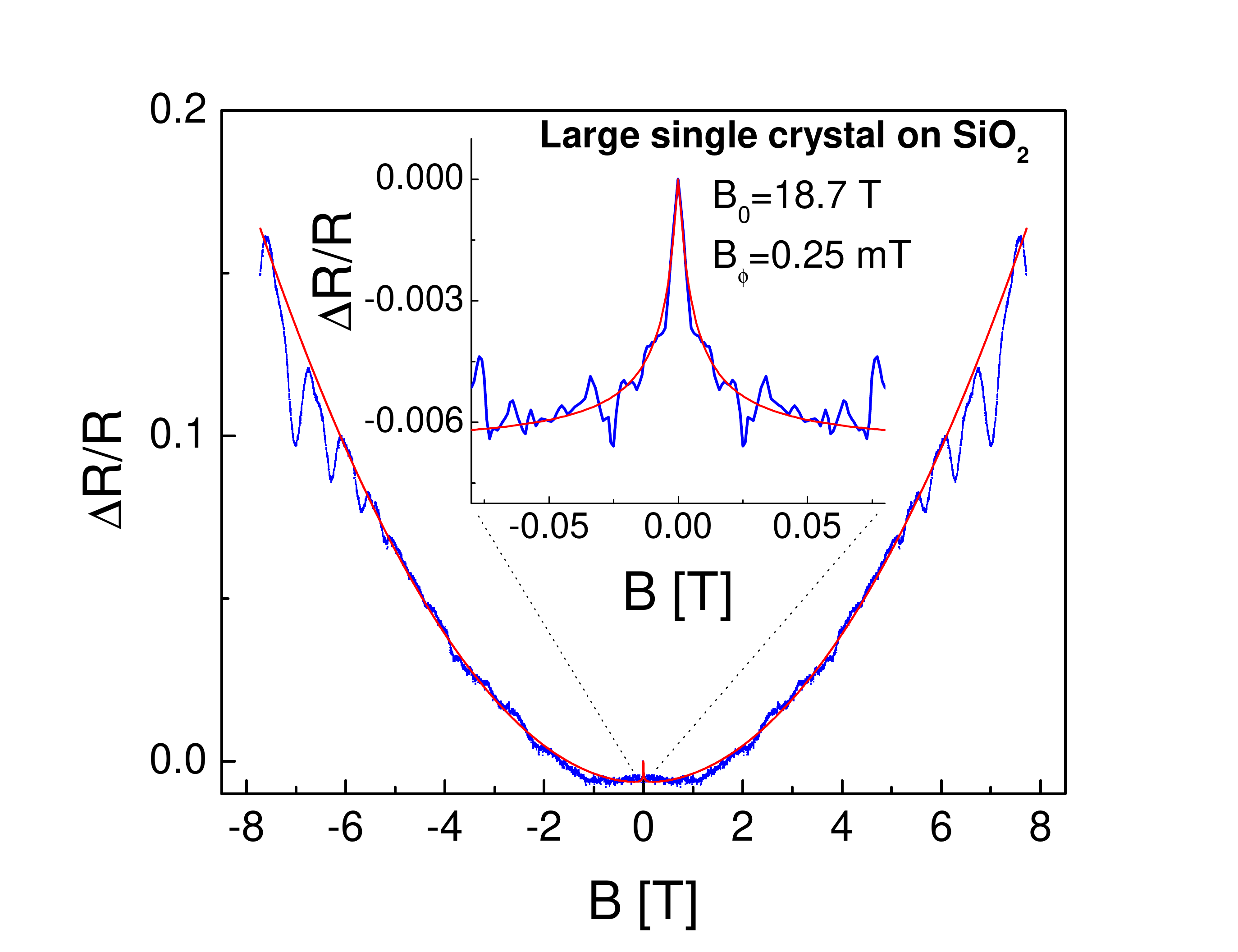}
\caption{The measured symmetrized magneto-resistance ($R_{xx}(B)+R_{xx}(-B)$) of a large scale graphene sample on glass measured at $V_G=+80V$ (blue dots). Shown is the relative resistance and the fit (red line) using equation (\ref{expfit}) with fitting parameters $B_{c\phi}=0.25$mT, $L_\phi/L_c=0.0067$, $B_0=18.7T$, and $B_e=3.5$mT}
\label{C13}
\end{figure}

We now turn to the higher mobility case, where we consider a graphene field effect device made out of a large ($\sim 200 \mu$m diameter) single graphene crystal grown by CVD. Because of its single crystal nature, this sample has a mobility comparable to exfoliated graphene on silicon oxide. From the gate voltage dependence we extracted a field effect mobility of about $\mu_n\simeq 0.43$m$^2$/Vs for electrons and $\mu_p\simeq 0.63$m$^2$/Vs for holes. The CNP was at $V_G=5.4V$. Using equation (\ref{expfit}) to fit the experimental data we can extract the various length scales and obtain $L_\phi\simeq 810$nm, $L_c\simeq 0.1$mm, $L_e\simeq 220$nm and $\sqrt{\mu_n\mu_p}\simeq 0.5$. For the median mobility, we used the experimental resistivity ratio of $\rho_{max}/\rho(+80V)\simeq 9$ and find a remarkable agreement to the field effect mobility extracted from the gate voltage dependence. For the residual density we obtain $n_0\simeq 5.8\times 10^{11}$cm$^{-2}$ using $\mu^{-1}=en_0\rho_{max}$.

The overall experimental behavior can be summarized as follows: at low temperatures, very low mobility samples show a very wide peak in the resistance at zero field (as shown in figure \ref{lowmu}). In this regime it is important to include strong localization effects to understand the width of the peak, since $L_c$ is comparable to $L_\phi$. The effective width is then determined by $\sim1/L_c^2+1/L_\phi^2$. With increasing mobility, the peak becomes increasingly sharper (figures 5-7) and the relevant length scale becomes $L_\phi$, who's inverse square determines the peak width. In addition to the WL effect (sharp negative magnetoresistance), which all samples show, there is often a parabolic positive magnetoresistance which is mainly due to the 2CT effect because of large scale inhomogeneities in addition to WAL in very clean samples.

\section{Conclusion:}
We presented a comprehensive picture of the observed magnetoresistance in graphene spanning all disorder levels. At high disorder, localization becomes the dominant mechanism, where the field dependence can be understood in terms of the field dependent localization length, which interestingly follows the field dependence of the coherent return probability and hence the field dependence of the the WL correction. In this picture, WL is simply a consequence of the field dependent localization length. Both, WL and strong localization, are due to short-range scattering. Long range scattering causes a positive magnetoresistance effect on top of the negative WL magnetoresistance, which stems from WAL (due to intravalley scattering) and from the presence of two types of carriers close to the CNP (the 2CT). While WL is generic to all disordered systems, both WAL and the 2CT are specific to graphene, where both play an important role in the understanding of magneto-transport.

\section{Acknowledgments:}
We thank NSERC and FQRNT for financial assistance.

\end{document}